\documentclass{amsart}
\usepackage{amsmath}
\usepackage{amsfonts}
\usepackage{amscd}

\let\eps=\epsilon

\newcommand{\bra}{\langle}
\newcommand{\ket}{\rangle}
\newcommand{\dq}{\dot{q}}
\newcommand{\dQ}{\dot{Q}}
\newcommand{\pder}[2]{\frac{\partial #1}{\partial #2}}
\newcommand{\I}{\mathcal{I}}
\newtheorem{prop}{Proposition}
\newtheorem{teo}{Theorem}

\newcommand{\R}{\ensuremath{\mathbb{R}}}
\newcommand{\pderd}[3]{\frac{\partial^2 #1}{\partial #2 \partial #3}}
\newcommand{\de}{\dot{\epsilon}}
\def\ni{\noindent}
\newcommand\ie{{\it i.e.}}

\newbox\Ancha
\def\gros#1{{\setbox\Ancha=\hbox{$#1$}
   \kern-.025em\copy\Ancha\kern-\wd\Ancha
   \kern.05em\copy\Ancha\kern-\wd\Ancha
   \kern-.025em\raise.0433em\box\Ancha}}
 3
 2
 1
\title{Variational equations of Lagrangian systems and
Hamilton's principle}

\author[H.N N\'u\~nez-Y\'epez]{H.\ N.\  N\'u\~nez-Y\'epez$^{(1)}$}
\thanks{$^{(1)}$ All authors contributed equally to this paper. Partially supported by PAPIIT-IN122498}
    \address{$^{(1)}$Departamento de F\'{\i}sica, 
        Universidad Aut\'onoma Metropolitana-Iztapalapa, 
         Apar\-tado Postal  55-534 Iztapalapa 09340 D.\ F., M\'exico.}
         \email{nyhn@xanum.uam.mx}

\author[J. Delgado]{Joaqu\'{\i}n Delgado$^{(2)}$}
\thanks{$^{(2)}$Partially supported by Conacyt grant 32167-E}
     \address{$^{(2)}$Departamento de Matem\'aticas, Universidad Aut\'onoma Metropolitana-Iztapala\-pa, 
	Apartado Postal  55-534, Iztapalapa 09340 D. F., M\'exico.}
	\email{jdf@xanum.uam.mx}

\author[A.L. Salas-Brito]{A.\ L.\ Salas-Brito$^{(3)}$}
\thanks{$^{(3)}$ Part of this work was done while visiting the Department of Physics, Emory University}
    \address{$^{(3)}$Laboratorio de Sistemas Din\'amicos, 
		Departamento de Ciencias B\'asicas, 
		Universidad Aut\'onoma Metropolitana-Azcapotzalco, 
		Apartado Postal 21-726, Co\-yoac\'an 04000 D.\ F., M\'exico.}
		\email{asb@correo.azc.uam.mx}

\date{\today}

\begin{document}

\begin{abstract}
\ni We  discuss a recently proposed variational principle for deriving the variational equations associated to any Lagrangian system. The principle gives
simultaneously  the Lagrange  and the  variational equations of the system.  We define a new  Lagrangian in an extended  configuration space ---which we call D'Alambert's--- comprising both the original coordinates and  the compatible ``virtual displacements'' joining two solutions of the original system. The variational principle is  Hamilton's with the new Lagrangian. We use this formulation to obtain  constants of motion 
in the Jacobi equations of any Lagrangian system with symmetries. These constants are related to constants in the original system and so with symmetries of the original Lagrangian. We cast our approach in an intrinsic coordinate free  formulation. Our results  can be of interest for reducing the dimensions of the equations that characterize perturbations in a Lagrangian control system.
\end{abstract}


\maketitle

\section{Introduction}

Extremum principles have a great importance in dynamics and in control \cite{1},\cite{Sus1}, \cite{Sus0}. They can be used to attain general formulations of the theory and can be of assistance in the solution of particular problems \cite{2},\cite{3}. The purpose of this contribution is to discuss  a recently proposed extremum principle which generates both the equations of motion  
and the so-called variational equations describing deviations between 
any two  solutions of  a Lagrangian system \cite{4}. The variational principle is Hamilton's but with a different Lagrangian.  Since we are dealing with Lagrangian dynamical systems, in which such ``least action'' principle is assumed to be valid, the variational equations describe deviations between extremal curves in a certain configuration space so they, slightly abusing the language, may be called the Jacobi variational equations  of the Lagrangian system as we have done in \cite{4}. Please note that though we use a 
notation with a strong flavour of classical mechanics \cite{5}, the formulation  is not limited in any way and can  be applied to non-linear  evolution equations, 
like  the  Einstein, the Korteweg-DeVries, or  the Kamdomtsev-Petviashvili equations \cite{6}--\cite{9}.  The  intrinsic formulation of  section 4  should make clear this matter. The variational equations can also be used to study the stability of  solutions and as starting points to evaluate their Liapunov spectrum as we have succintly discussed in \cite{4}.  They are, besides, describable in the vertical extension of the Lagrangian formalism \cite{10}. We should mention previous related  work  along the same lines \cite{11} and an older contribution  with different purposes \cite{12}.  

The article is organized as follows: In section 2 we define the new  Lagrangian, use it to formulate the variational principle whose extremals are solutions of the Lagrange  and the variational equations of the system. In this section we also  discuss basic features of the formulation. In section 3, we address the existence of constants of motion in the variational equations and its relation with constants in the original  system. In section 4 we cast our approach in an intrinsic language, making connection with the  symplectic manifold that can be associated with the prolongation of the original Lagrangian. If you are  interested mainly in the mathematical setting of the formulation this is the section to read. In section 5 we give various examples in which the approach may be used. 

\section{The variational principle}

Let us consider then a $N$-degree of freedom  dynamical system endowed with a Lagrangian function $L(q_a \dot q_a, t), \; a=1\dots N;$ defined in the tangent bundle $TQ$ of its configuration space $Q$, \ie\ $ L: TQ\to \R$. From the Lagrangian we usually construct the action \cite{2},\cite{3}

\begin{equation}
S[q(t)]= \int_{t_1}^{t_2} L(q_a \dot q_a)\, dt  \label{1} 
\end{equation}

\noindent where we are using square brackets to indicate that $S$ is a functional of the paths $q_a(t)$ joining two given points, $q_a(t_1)$ and $q_a(t_2)$ ($a=1,\dots,N$), in $Q$  at two fixed instants of time $t_1$ and $t_2$.  The extremalization of $S$ directly leads to the Lagrangian equations of motion of the system [equations (\ref{6}) below]. 
Let us now consider a different configuration space $D$, comprising both the original configurations of the system plus all the possible ``virtual displacements'' joining any two extremal paths of (\ref{1}).  With the help of $L$,  we can  define a new (the mathematical meaning of all this should become clear in sections 4.2 and 4.3) Lagrangian $\gamma({\bf q}, {\bf\dot q}, {\gros \epsilon}, \dot {\gros\epsilon}, t)$ \cite{4} as

\begin{equation}
\gamma({\bf q}, {\bf\dot q}, {\gros \epsilon}, \dot {\gros\epsilon}, t)\equiv 
\frac{\partial L}{ \partial {\dot q}_a} \dot \epsilon_a + \frac{\partial L}{ \partial 
q_a} \epsilon_a \label{2} 
\end{equation}

\ni here, as in all of the paper,  the summation convention  is implied for repeated indices. The function $\gamma: TD \to \R$, where $TD$ is the tangent bundle of $D$, is central in our formulation. The $2N$-dimensional D'Alambert configuration space $D$ is assumed, mostly in sections 2 and 3,  to be coordinatized by $(q_a, \epsilon_a),\; a=1,\dots, N$. The $N$-component object ${\gros \epsilon} =(\epsilon_1,\epsilon_2,\dots, \epsilon_N)$ stands for the  displacement from an extremal path of (\ref{1}) to another, and  $ {\bf \dot {\gros \epsilon}} = (\dot\epsilon_1,\dot\epsilon_2,\dots, \dot\epsilon_N)$ for its corresponding velocity. This means that ${\gros\epsilon}$ plays the role of the variational field associated with trajectories of the original system: ${\bf q}'= {\bf q}+ {\gros \epsilon}$ and ${\bf \dot q}'= {\bf \dot q} + {\dot \gros \epsilon}$, where both $\bf q$ and $\bf q'$ are extremal paths \cite{1}, \cite{2}, \cite{4} of the original action (\ref{1}).  A useful property of $\gamma({\bf q}, {\bf \dot q},{\gros \epsilon}, {\dot \gros \epsilon}, t)$ is that it  is an explicit function of time {\sl only} when  $L$ is non-autonomous.   

The function $\gamma$ is used to define the new (``displaced'') action functional 

\begin{equation}
 \Sigma[{\bf q}(t), {\gros\epsilon}(t)]=\int_{t_1}^{t_2}
 \gamma({\bf q}, {\bf \dot q}, {\gros\epsilon}, {\bf \dot \gros\epsilon},t)  
dt \label{3}
\end{equation}

\ni of the paths joining two given configurations $(q_1,\epsilon_1)$ and$(q_2,\epsilon_2)$ 
of the varied system between two fixed instants of time $t_1$ and $t_2$. We are calling ``time'' the parameter appearing in (\ref{3}), but it can be any other useful parameter, for example, the arc length, or even may be any finite {\sl set} of parameters. The statement of the variational principle is just Hamilton's, that is, $\Sigma$ is extremal when the system follows its actual path in $D$\cite{4}

\begin{equation}
\delta \,\Sigma[{\bf q}(t), {\gros\epsilon}(t)]=0. \label{4}  
\end{equation}

\ni The extremalization is done  varying the path but mantaining the endpoints 
and the time fixed.  The conditions for the functional $\Sigma[q(t), \epsilon(t)]$ to 
be an extremum  are the $2N$ Euler-Lagrange equations \cite{1},\cite{2},\cite{12}

\begin{equation}
 \frac{d}{dt}\left( \frac{\partial \gamma}{ \partial \dot \epsilon_a} 
\right)-\frac{\partial \gamma}{ \partial \epsilon_a}=0,\quad
\frac{d}{ dt }\left( \frac{\partial \gamma}{ \partial \dot q_a} \right)-
\frac{\partial \gamma}{ \partial q_a}=0,\qquad a=1,\dots,N; \label{5}
\end{equation}

\ni or, using the definition (\ref{2}) in the preceding equations, we obtain  the  Lagrange equations of the original system

\begin{equation}
\frac{d}{ dt }\left( \frac{\partial L}{ \partial \dot q_a} \right)-
\frac{\partial L}{ \partial q_a}=0,\quad a=1,\dots, N, \label{6}
\end{equation}

\noindent plus the  linear variational equations:

\begin{equation} 
  M_{ab}\ddot \epsilon_b + C_{ab}\dot \epsilon_b  + K_{ab}\epsilon_b=0,   
  \quad a =1,\dots, N. \label{7}                 
\end{equation}
 
\noindent Equations (\ref{7}) describe the evolution of the deviation, 
${\gros \epsilon}$, of a varied trajectory from an 
unperturbed one. The $N\times N$ matrices $M$, $C$ and $K$, are 

\begin{eqnarray}
 M_{ab}   &=&\left(\frac{\partial^2 L}{ \partial \dot q_a \partial \dot
q_b}\right), \\ 
C_{ab} &=&\left[\frac{d}{ dt} 
   \left( \frac{\partial^2L}{ \partial\dot q_a \partial\dot q_b}
      \right) + \frac{\partial^2 L}{ \partial \dot q_a \partial q_b} -
	\frac{\partial^2 L}{ \partial \dot q_b \partial q_a}\right], \\
K_{ab}&=&\left[\frac{d}{
dt}\left( \frac{\partial^2L}{ \partial\dot q_a \partial q_b} \right)  -
\frac{\partial^2 L}{ \partial 
 q_a \partial q_b}\right], \quad
a,b=1,\dots,N. 
\end{eqnarray}

\noindent It shoud be clear that equations (\ref{7}) are the  variational equations of the original system \cite{1}, \cite{4}, \cite{6}. One of the interesting properties of these equations is that the matrices $M$, $C$, and $K$, and, hence,  the variational equations (\ref{7}), may be regarded as  time-independent as long as $L$  is autonomous. This property, that does not happen in the standard description, may have important consequences in analysing perturbations to optimal control problems \cite{Sus0}\cite{VdS} \cite{Nij}. At this point, it is worthwhile to emphasize four important features: 

\begin{enumerate}

\item  The function $\gamma : TD\to \R$ plays the role of a new Lagrangian describing the original  system plus its response to  ``virtual displacements'' or perturbations.  It therefore is also useful for studying stability \cite{4}.

\item Equations (\ref{6}) and (\ref{7}) are invariant under arbitrary point transformations (\ie\  changes of coordinates in $Q$) : $Q_a=f_a(q,t),\; a=1,\dots, N$, where the  $f_a$ are   functions, $f_a: D\to D$, assumed to be  invertible (\ie\ $\det(df(q))\neq0,\, q\in D$) and at least $C^{1}$.  As happens in any Lagrangian description, this result is proved by the very existence of the variational principle (\ref{3}); \cite{1},\cite{2},\cite{14}. 

\item The $\epsilon$-derivatives of $\gamma$ reduce to corresponding $q$-derivatives of $L$

\begin{equation} 
\frac{\partial \gamma} {\partial\dot\epsilon_a}= 
\frac{\partial L} {\partial \dot q_a},\qquad \hbox{and} 
\qquad \frac{\partial  \gamma} {\partial  \epsilon_a}= 
\frac{\partial L} {\partial  q_a}; \label{9}
\end{equation}

\item As a consequence of the previous property, $\gamma$ is 
a  first-order homogeneous function of the virtual displacements $\epsilon_a$ 
and   velocities $\dot \epsilon_a$
\begin{equation}
\gamma= \frac{\partial \gamma} {\partial \epsilon_a}\epsilon_a +
 \frac{\partial \gamma} {\partial \dot\epsilon_a}\dot\epsilon_a. \label{10}
\end{equation}
\end{enumerate}

\section{Symmetries and constants of motion}

One of the most important consequences of any Lagrangian description 
using a variational principle, is the close association between the symmetries of the Lagrangian  and the existence of constants of motion  \cite{15}, that is, functions that are conserved along integral curves of the Lagrangian vector field. There is a control-theory version of such result, saying that every symmetry of a control system gives rise to a conservation law along biextremals \cite{Sus0}. This theorem makes the following results   important for perturbed control systems.
  
The association between symmetries and conservation laws is  shared by the the function $\gamma$. The precise sense  is the following: Any symmetry of $L$ can be carried over to a symmetry of $\gamma$, where they imply the existence of constants of motion in the variational equations. Although  a general proof can be given, we prefer to illustrate this result in what follows ---though our discussion of what we have called inherited constants comes close to being an informal proof.
   
\subsection{Invariance under time translations} 

If the Lagrangian of the system is autonomous,  then  there exist  a  well-known constant of motion \cite{1}, \cite{2},

\begin{equation}
H = \frac{\partial L} {\partial \dot q_b}\dot q_b -L. \label{11} 
\end{equation}

\ni $H$ can, in certain instances, be identified with the energy of the system.

In such autonomous case, $\gamma$ is autonomous too and the variational equations thus admit an analogous constant of motion ---which can be derived in an strictly similar fashion to $H$ \cite{1},\cite{2},\cite{4}--- namely

\begin{equation}
h = \frac{\partial \gamma} {\partial \dot q_b}\dot q_b +
\frac{\partial \gamma} {\partial \dot \epsilon_b}\dot \epsilon_b-\gamma.
\label{12} 
\end{equation}

\ni Using definition (\ref{2}), the constant $h$ can be recasted as \cite{4}

\begin{equation}
h = \frac{\partial \gamma} {\partial \dot q_b}\dot q_b -
\frac{\partial \gamma} {\partial \epsilon_b} \epsilon_b, \label{13}
\end{equation}

\ni or, using (\ref{11}), as

\begin{equation}
h=\frac{\partial H} {\partial \dot q_b}\dot \epsilon_b +
\frac{\partial H} {\partial  q_b} \epsilon_b. \label{14} 
\end{equation}

If we interpret $H$ as the energy, a  related interpretation for $h$ is that it is the first-order energy change in going from a solution to a displaced nearby one.

\subsection{Invariance under space translations}

The invariance of a system under space translations, the property frequently called by physicists the homogeneity of space, usually manifest itself in the independence of the Lagrangian on certain coordinates. When the Lagrangian does not depend on a specific coordinate $q_s$,  the coordinate is said to be {\sl ignorable} (respect to 
the original Lagrangian $L$), then its conjugate momentum is conserved. That is, if

\begin{equation}
\frac{\partial L} {\partial q_s}=0, \quad \hbox{then}\quad p_s\equiv 
\frac{\partial \gamma} {\partial\dot \epsilon_s}= \frac{\partial L} 
{\partial\dot q_s} \quad\hbox{is a constant.}  \label{15} 
\end{equation}

If $q_s$ is ignorable in $L$, then, as shown  in equation (\ref{15}), it is also ignorable in $\gamma$,  and $\epsilon_s$ is ignorable too, so  the momentum $p_s$ ---conjugated 
to $\epsilon_a$ in $\gamma$--- is conserved (as it should be since it is conserved in the original system!). 
Furthermore, $q_s$ is also ignorable in $\gamma$, hence the quantity 
(the momentum, $\pi_s$, conjugated to $q_s$ in $\gamma$)
\begin{equation}
\pi_s\equiv \frac{\partial \gamma} {\partial \dot q_s}=
\frac{\partial^2L} {\partial \dot q_s \partial \dot q_b} \dot\epsilon_b + 
\frac{\partial^2 L} {\partial\dot q_s \partial q_b}\epsilon_b, \label{16}
\end{equation}
is also a constant of motion in the variational equations.

After having proved the existence of the constants $h$  (equation \ref{12}) and $\pi$ 
(equation \ref{16}), it is worth formulating a more general result. As we show next such theorem follows from the close relationship between solutions to equations (\ref{6}) and constants of motion in the variational equations (\ref{7}). We call such first integrals, {\sl inherited constants of motion}.

\subsection{Inherited constants of motion}

Let us consider any constant of motion, $J({\bf q}, \dot{\bf q})$, of the original 
set of equations  (6). If we evaluate it on two nearby solutions of (6), ${\bf q}$ 
and ${\bf q'}={\bf q}+ {\gros \epsilon}$, separated by the Jacobi field 
${\gros \epsilon}$, the difference, $j({\gros\epsilon}, \dot{\gros\epsilon})\equiv 
J({\bf q}',\dot{\bf q}')-J({\bf q},\dot{\bf q})$, between these constant quantities 
is also trivially a constant, 

\begin{equation} 
\frac{d j({\gros \epsilon},\dot{\gros\epsilon})} {dt}=\frac{d} {dt}
\left[ J({\bf q}',\dot{\bf q}')-J({\bf q},\dot{\bf q}) \right] =0. \label{17} 
\end{equation}

\ni  The  constant, $j[{\gros \epsilon}]$, can be   expressed  as 
 
\begin{equation} 
j[{\gros \epsilon}] =  \left(\frac{\partial J} {\partial q_a}\epsilon_a+ \frac{\partial 
J} {\partial \dot q_a}\dot\epsilon_a \right). \label{18}
\end{equation}

\noindent This is precisely the form of $h$ in equation (\ref{14}).

Equation (\ref{18}) tells us how, given both a solution of equations (\ref{6}) and any one of its constants of motion, we can obtain a constant of motion in equations (\ref{7}). Equation (\ref{18}) can be directly proved to be a constant by computing its time derivative.  This result  establishes  a direct relationship of constants in the variational equations, like $j$, to constants in the original Lagrangian.  Related  results  are discussed in \cite{5},\cite{17}. 

 Furthermore, as in some non-linear evolution equations the constants of motion, $J[q(t)]$, are functionals (and not functions) of the solutions of  (\ref{6}), we need to pinpoint that  the constant  $j[\epsilon(t)]$  becomes a functional of the Jacobi 
fields $\epsilon(t)$, which must be given thus by

\begin{equation}
j[\epsilon(t)]=\int \frac{\delta J[q(t)]}{ \delta q(\tau)} \epsilon(\tau)\; d\tau, 
\label{19}   
\end{equation}

\noindent where ${\delta J[q(t)]/ \delta q(\tau)}$ stands for the functional derivative of  the constant functional $J[q(t)]$ \cite{16}\cite{riesz}. 

Having stablished the existence of the inherited constants (\ref{18}) [or (\ref{19})], 
it should be clear that any symmetry of the original Lagrangian is reflected in 
the existence of another constant in the Jacobi variational equations. Thus, 
Noether's theorem  also holds for the variational equations and can be used to reduce the dimensions of the variational system \cite{4};\cite{Sus0}. Related  
results  are discussed in \cite[section 111]{5}; \cite{9}.

\section{Intrinsic formulation}

In this section we formulate in an intrinsic manner our variational approach. First we consider an arbitrary  vector field $Y$ (at least $C^1$) defined on a manifold $M$ and recall the constructions of the {\em variational vector field\/} $T(Y)$ defined on the tangent bundle $TM$ and the {\em adjoint variational vector field\/}  $T^*(Y)$ defined on the cotangent bundle $T^*M$. As it is well known \cite{Sus1}, \cite{VdS},\cite{Nij},\cite{18},\cite{Mar},  $T^*(YX)$ posseses a natural structure:
it is a Hamiltonian vector field with respect to the the canonical  symplectic form 
on $T^*M$. On the other hand the variational vector field $T(Y)$ does not have a natural Lagrangian structure due to the fact that the Hamiltonian for $T^*(Y)$ is linear in the momenta and so there is no  natural Legendre transform $TM\to T^*M$ defined. 

Our main interest is in the case of $M=TQ$, {\it i.e.\/} the tangent bundle of a configuration space, and $Y=Y_L$,  a Lagrangian vector field on $TQ$ for the function $L\colon TQ\to \R$. We recall that if $L$ is not degenerate, the pull back of the canonical symplectic form $\omega_0$ in $T^*Q$ under the Legendre transformation $\mathcal{L}(L)\colon TQ\to T^* Q$, $\omega_L= \mathcal{L}^*\omega_0$, makes $(TQ,\omega_L)$ into a symplectic manifold and the Lagrangian vector field $Y_L$ is Hamiltonian for the energy function $H\colon TQ\to\R$. Also $Y_L$ is a second order differential equation: $\tau_{Q_{*}} Y_L  = Y_L\circ\tau$ 
where $\tau_Q\colon TQ\to Q$ is the  projection.

From what  was said in the first paragraph of this section, the adjoint variational vector field $T^*(Y_L)$,  defined on $T^*(TQ)$, has a natural Hamiltonian form. We want to dilucidate what the structure of the variational vector field $T(Y_L)$, defined on $T(TQ)$, is. A naive guess is that $T(Y_L)$ should be Lagrangian and  the Lagrangian should be the prolongation \cite{VdS},\cite{Sus1} $\dot{L}\colon T(TQ)\to \R$ of $L\colon TQ\to\R$: $\dot{L}(\xi) = dL_{\pi(\xi)}(\xi)$,
for $\xi\in T(TQ)$.  A moment of caveat shows that this cannot be true. In fact,
taking coordinates $(q,\dot{q},\epsilon,\de)$ for $T(TQ)$, and
$(q,\dot{q},p,\pi)$ in $T^*(TQ)$, the prolongation is given by

\begin{equation}\label{sim}
\frac{dL}{dt}(q,\dot{q},\epsilon,\de) = 
\pder{L}{q}(q,\dot{q})\epsilon+\pder{L}{\dot{q}}(q,\dot{q})\de
\end{equation}

\noindent and its Legendre transform $T(TQ) \xrightarrow {\mathcal{L}(\dot{L})} T^*(TQ)$,
 is given by
\begin{eqnarray}
p &=& \pder{\dot{L}}{\epsilon} = \pder{L}{q},\\
\pi &= & \pder{\dot{L}}{\de} = \pder{L}{\dot{q}}
\end{eqnarray}

\noindent but then the pull back of $\theta_0=p\,dq+\pi\,d\dot{q}$ is

\begin{equation}
\theta_{\dot{L}} = \mathcal{L}(\dot{L})^*\theta_0 = 
(p\circ \dot{L})\,dq + (\pi\circ \dot{L})\,d\dot{q} = \pder{L}{q}\,dq+\pder{L}{\dot{q}}\,d\dot{q}
\end{equation}

\noindent that is $ \theta_{\dot{L}} = dL \circ \tau_{TQ}$, there $\tau_{TQ}\colon T(TQ)\to TQ$
is the projection. This in turn implies $d \theta_{\dot{L}}\equiv 0$ (!). Thence the interpretation is untenable.

\subsection{The variational and adjoint variational vector fields}

For setting straight the formalism, let us  recall briefly the appropriate construction: Let $M$ be a manifold   and let $\tau_M:TM\to M$, $\tau^*_{M}\colon T^*M\to M$ be the tangent and cotangent bundles, respectively. Given a   vector field on $M$, that is  a section $Y\colon M \to TM$, the variational vector field associated to $Y$,
herein denoted by $T(Y)$, is defined as follows: Given  $m\in M$ there exists a neighborhood $B$ of $m$ and $\epsilon>0$ such that the local flow
$\Phi_t \colon B_{\delta}(m)\to M$ is defined for $t\in  (-\epsilon,\epsilon)$.
The lift $T(\Phi_t): TM|_{\pi^{-1}(B)}\colon \pi^{-1}(B)\to TM$ 
then defines a local flow and its infinitesimal generator is the {\em variational 
vector field } $T(Y)$. Similarly, the lifting to the cotangent bundle
$T^{*}(\phi_t): TM|_{p^{-1}(B)}\colon \pi^{-1}(B)\to T^*M$ defines a local flow and its 
infinitesimal generator is called the {\em adjoint variational vector field} $T^*(Y)$.

The names are well suited since in local coordinates, $(x,v)$ for $TM$ and $(x,p)$ 
in $T^*M$ (we think $x,v$ as column vectors; $p$ a row vector), the variational 
and adjoint variational  vector fields are 
given respectively as

\begin{subequations}\label{var}

	\begin{align*}
 	\dot{x} &= f(x),         &  \dot{x} &= f(x),    \nonumber \\
	\dot{v} &= Df(x)v;\qquad   (\theequation.a)  &  \dot{p} &= -p Df(x)  \qquad (\theequation.b)
     \end{align*} 
\end{subequations}

Actually the adjoint variational vector field (\ref{var}.b) 
admits the  Hamiltonian form 
\begin{align}
\dot{x} &= \pder{H}{x},\\
\dot{p} &= -\pder{H}{x}.
\end{align}

\noindent where 
\begin{equation}\label{hamiltonian}
H(x,p) = p f(x).
\end{equation}

\noindent This form is compatible with the standard symplectic form on 
$T^* M$. As we mentioned earlier in this section,
 there is no natural Lagrangian structure for (\ref{var}.a) since 
the Legendre transformation $(x,p) \xrightarrow{\mathcal{L}^{-1}(H)}(x,v)$ is
not defined  due to the linear dependence of $H$ on the momenta (\ref{hamiltonian}).

\subsection{Prolongation of a system}

There is a related concept of extending a given system  known as its {\em prolongation}
\cite{Sus0}, \cite{VdS}, \cite{Nij}.
The motivation comes from the  following argument: if the system is given by  $\dot{x} = f(x)$,
then  by taking  time derivatives along solutions one gets

\begin{equation}
\ddot{x} = Df(x)\dot{x} =  Df(x)f(x),\,
\dddot{x} = D^2f(x)\cdot(Df(x),f(x) + Df(x)\cdot (Df(x)\cdot f(x)),\, etc.
\end{equation}

\noindent Considering the first prolongation $\ddot{x} = Df(x)f(x)$, it
can be written as the first order system
\begin{equation} \label{first}
\begin{array}{rcl}
\dot{x} &=& v,\\
\dot{v} &=& Df(x)\cdot f(x).
\end{array}
\end{equation}

\noindent A natural question is if this system admits a Lagrangian structure. 

\begin{prop} Consider the system $\dot{x}=f(x)$ and suppose that 
$\pder{f_i}{x_j}=\pder{f_j}{x_i}$, that is  $Df(x)$ be symmetric. 
 Then the first prolongation of the system
 (\ref{first}) is Lagrangian for the  function
\begin{equation}
L(x,v) = \frac{1}{2}|v|^2 + \frac{1}{2}|f(x)|.
\end{equation}
\end{prop}
The proof is elementary and will be omited.
\qed

\bigskip
The symmetry condition can be rephrased in terms of the graph

\begin{equation}
G_f = \left\{(x,f(x)^T)\mid x\in \mbox{Dom}\, f \right\}\subset R^n\times (R^n)^*
\end{equation}

\noindent as being Lagrangian \cite{AM}, \cite{VdS} (here the notation $f(x)^T$ means simply that $f(x)$ is viewed as a covector).

If $f\colon M\to N$,  its  prolongation  \cite{Sus0}\cite{VdS} is the map 
$\hat{f}\colon T(M)\to T^*(N)$ defined as follows: For $\xi\in T_{x}M$ and 
$\alpha\in T_{f(x)}^* (M)$, let $\bra\, ,\ket$ denote the natural pairing, then

\begin{equation}
\hat{f}(\xi)(\alpha)  = \bra f_{*x}(\xi),\alpha\ket.
\end{equation}

\noindent In local coordinates $(x,v)\in\R^n\times\R^n$ for $TM$ and $(y,p)\in\R^n\times(\R^n)^*$
for $T^*(N)$ the first prolongation is given by

\begin{align}
y&=f(x),\\
p &= v^T Df(x)^T.
\end{align}

\ni In the case of a function $f\colon M\to \R$, $T_{f(x)}^*\R \simeq \R$ and
the prolongation can be identified with the differential of the map $df\colon TM\to\R$

\begin{equation}
\hat{f}(\xi_x) = df_{x}(\xi_x).
\end{equation}

\subsection{Variational equations of Lagrangian systems}

We now consider the case of $M=TQ$ for some configuration space $Q$, and 
$L\colon TQ\to\R$ a non degenerate Lagrangian for the vector field $Y$ on $TQ$. The prolongation is then  the a  map $\dot{L}\colon T(TQ)\to\R$.  
As was mentioned at the beginning of this section, the variational vector field
$T(Y)$ cannot be viewed as the Lagrangian vector field for the prolongation $\dot{L}$
in an obvious way since the pullback of the canonical $1$-form in $T^*(TQ)$ to $T(TQ)$
by the Legendre transform $\mathcal{L}(\dot{L})$ of the prolongation $\dot{L}$
is an exact differential.

Instead we consider the {\em D'Alambert phase space} consisting of pairs $(q,\epsilon)$
where $q$ is a given configuration and $\epsilon$ is a ``virtual displacement''. 
More formally we consider a subbundle , $D\subseteq  Q$
then $TD$ an be embeded in $T(TQ)$ under the map  $(q,\epsilon,\dot{q},\de)
\xrightarrow{\alpha}(q,\dot{q},\epsilon,\de)$. Now 
view the prolongation as a map $\gamma\colon TD\to\R$, that is

\begin{equation}\label{gamma}
\gamma(q,\eps,\dot{q},\de) = \pder{L}{q}(q,\dot{q})\eps + 
\pder{L}{\dot{q}}(q,\dot{q})\de
\end{equation}

\ni (in fact $TD$ is the natural domain of definition of the prolongation, since the ``virtual displacements'' are defined by choosing the subbundle).

Notice that the embeding $\alpha$ is well defined since the $\eps$'s are elements
of the tangent space $T_qQ$, and that $D$ is not necessarily a tangent bundle, that is,
it is not necessarily an integrable distribution. Thus we can consider even
holonomic or nonholonomic constraints in the specification of $D$.

\begin{teo} Let $\omega_{\gamma} = \mathcal{L}(\dot{L})^*\omega_0$ be the pullback of the
canonical symplectic form on $T^*D$ under the Legendre transformation of the prolongation
of $L\colon TQ\to\R$,  viewed as a map $\gamma \colon TD \to\R$. Then $T(Y_L)$ viewed as a vector field on $TD$ is a Lagrangian vector field for the energy function $h\colon TD\to\R$. In local coordinates $(q,\eps,\dot{q},\de)$ for $TD$,

\begin{equation}
h(q,\eps,\dot{q},\de) = \pder{\gamma}{\dot{q}}\dot{q}-\pder{\gamma}{\eps}\eps,
\end{equation}
\ni which coincides with (\ref{14}).
\end{teo}

\proof We carry on the proof in local coordinates. Let $(q,\eps,\dot{q},\de)$ be
coordinates in $TD$ and $(q,\eps,\pi,p)$ coordinates in $T^*D$.
The canonical $1$-form in $T^*D$ is given by $\theta_0 = \pi\,dq+p\,d\eps$. The Legendre
transform $\mathcal{L}(\gamma)$ is given by

\begin{align}
\pi &= \pder{\gamma}{\dot{q}},\\
p &= \pder{\gamma}{\de} = \pder{L}{\dot{q}}.
\end{align}

\ni then

\begin{equation}
\theta_{\gamma} =\pder{\gamma}{\dot{q}}\,d\,q + \pder{L}{\dot{q}}\,d\eps
\end{equation}

An straightforward computation shows that

\begin{equation}
\begin{split}
\omega_{\gamma} = ~{} &\pderd{\gamma}{q_k}{\dq_j}dq_k\wedge dq_j +
\pderd{\gamma}{\dq_k}{\dq_j}d\dq_k\wedge dq_j + 
\pderd{\gamma}{\eps_k}{\dq_j}d\eps_k\wedge dq_j \\
+ 
&\pderd{\gamma}{\de_k}{\dq_j}d\de_k\wedge dq_j +
\pderd{\gamma}{q_k}{\de_j} dq_k\wedge d\eps_j  +
\pderd{\gamma} {\dq_k}{\de_j} d\dq_k\wedge d\eps_j.
\end{split}
\end{equation}

\ni Consider the second order equation (or spray)

\begin{equation}
T(Y_L) = \dq\pder{}{q}+\de\pder{}{\eps}+ C\pder{}{\dq}+D \pder{}{\de}
\end{equation}

\ni then

\begin{equation}
\begin{split}
T(Y_L)\lrcorner\omega_{\gamma} &= \pderd{\gamma}{q_k}{\dq_j}(\dq_k dq_j-\dq_jdq_k)
  -\pderd{\gamma}{\dq_k}{\dq_j}\dq_jd\dq_k 
   -\pderd{\gamma}{\eps_k}{\dq_j}\dq_j d\eps_k 
  -\pderd{\gamma}{\de_k}{\dq_j}\dq_jd\de_k \\
& + \pderd{\gamma}{q_k}{\de_j}\dq_kd\eps_j +
  \pderd{\gamma}{\eps_k}{\dq_j}\de_k dq_j 
  -\pderd{\gamma}{q_k}{\de_j}\de_jdq_k 
  -\pderd{\gamma}{\dq_k}{\de_j}\de_jd\dq_k\\
& + \pderd{\gamma}{\dq_k}{\dq_j}C_k dq_j + 
  \pderd{\gamma}{\dq_k}{\de_j}  C_k d\eps_j +
 \pderd{\gamma}{\de_k}{\dq_j} D_k dq_j.
\end{split}
\end{equation}

\ni The differential of $h$ is

\begin{equation}
\begin{split}
dh &= \left( \pderd{\gamma}{q_k}{\dq_j}dq_k + \pderd{\gamma}{\dq_k}{\dq_j}d\dq_k
 + \pderd{\gamma}{\eps_k}{\dq_j} d\eps_k + \pderd{\gamma}{\de_k}{\dq_j}d\de_k
  \right) \dq_j \\
& + \pder{\gamma}{\dq_j}d\dq_j -
 \left(\pderd{\gamma}{q_k}{\eps_j}dq_k + \pderd{\gamma}{\dq_k}{\eps_j}d\dq_k
   \right)\eps_j - \pder{\gamma}{\eps_j}d\eps_j.
\end{split}
\end{equation}

\ni Equating coefficients of $d\eps_k$ in the equation 
$T(Y_L)\llcorner\omega_{\gamma}=-dh$ 
one gets

\begin{equation}
\pder{}{q_j}\left(\pder{L}{\dq_k}\right)\dq_j +
\pder{}{\dq_j}\left(\pder{L}{\dq_k}\right)C_j = \pder{L}{q_k}
\end{equation}

\ni by using the fact that $C_k= \ddot{q}_k$ (here the dots means  derivative with
respect to time) wer recover the orignal Lagrange equations

\begin{equation}
\frac{d}{dt}\left(\pder{L}{\dq_k}\right) = \pder{L}{q_k}.
\end{equation}

\ni Equating coefficients of $d\de_k$ and $d\dq_k$, sendous  identities are obtained. 
Finally equating the coefficients of $dq_k$ lead to the linearized equations (\ref{7}).
\qed

\section{Examples }
A few examples will be presented where the above  formulation can be
applied.

\bigskip
\noindent{\bf Geodesic flow}. Here the Lagrangian is just the square of the
length 

\begin{equation} 
L=\frac{ 1}{2}g_{ab}\dq_a\dq_b
\end{equation}

\ni where  $g_{ab}$ is the first fundamental form of the metric. The linearized equations
are properly known as {\em Jacobi equations} or as equations of {\em  geodesic displacement}. The prolonged Lagrangian is

\begin{equation}
\gamma = \pder{g_{ab}}{q_k}\eps_k \dq_a\dq_b + g_{ab}\dq_a\de_b.
\end{equation}

The Lagrange equations give the geodesic equations

\begin{equation}
\ddot q_a+ \Gamma_{abc}\dot q_b \dot q_c=0,
\end{equation}

\ni and the equations of geodesic deviation

\begin{equation}
\ddot \eps_a= R_{abcd}\dot q_b \dot q_d\,\eps_c, 
\end{equation}

\ni where, $\Gamma_{abc}$ is the affine connection or Christoffel's symbol,  $R_{abcd}$ is the Riemann tensor, and  the dots stand in this case for derivatives respect the arc length $s$ \cite{6}\cite{wein}.

\bigskip
\noindent{\bf Newton's equations.} Here  

\begin{equation}
 L= \frac{ 1}{ 2}m_{ab}\dq_a\dq_b-V(q),
\end{equation}

\ni where $m_{ab}$ is a constant  symmetric matrix.
The prolonged Lagrangian is

\begin{equation}
\gamma = -\pder{V}{q_a}\eps_a + m_{ab}\dq_a\de_b.
\end{equation}

\ni The Lagrange equation

\begin{equation}
\frac{d}{dt}\left(\pder{\gamma}{\de_k} \right) = \pder{\gamma}{\eps_k},
\end{equation}

\ni leads to

\begin{equation}
m_{ak}\ddot{q}_k = -\pder{V}{q_k}, 
\end{equation}

\ni which are Newton's equations. In the case of the solution being an equilibrium point
with the configuration a minimum of the potential, 
these are the equations of small oscillations.
Lagrange equation
\begin{equation}
\frac{d}{dt}\left(\pder{\gamma}{\dq_k} \right) = \pder{\gamma}{q_k}
\end{equation}

\ni then becomes

\begin{equation}
m_{ak}\ddot{\eps}_k + K_{ak}\eps_k=0, 
\end{equation}

\ni where 

\begin{equation}
K_{ab} = \pderd{V}{q_a}{q_b}.
\end{equation}

\bigskip
\noindent{\bf  Plane motions in an arbitrary 2-potential.}

Choose a reference curve and let $s=$ arc length,  $z=$ normal distance from
the curve, and  $\rho=$
radius of curvature; then taking $(s,z)$ as generalized coordinates  the 
Lagrangian is  \cite{4,12,Chandra}

\begin{equation}
L = \frac{ 1}{ 2}\left[\dot{z}^2 + \frac{\dot{s}^2}{\rho^2}(\rho+z)\right]
-V(z,s).
\end{equation}

\ni The linearized equations obtained by computing the prolonged Lagrangian yields

\begin{equation}\label{sta}
\ddot{\eps}_z + \frac{3}{\rho}\left(\frac{\dot{s}^2}{\rho}+\frac{\rho}{3}
\frac{\partial^2 V}{\partial z^2}\right)\eps_z= \frac{2h}{\rho}.
\end{equation}

\ni Here is worth to say that {\em we choose\/} the ``virtual displacements'' $\eps_s$ and
$\eps_z$ to be planar ones.
The constant of motion $h$ can be expressed as 

\begin{equation}
h  =\dot{s}\dot{z}\left(\frac{\dot{s}^2}{\rho}+\pder{V}{z}\right)\eps_z+
\pder{V}{s}\eps_s.
\end{equation}

We  pinpoint that from equation (\ref{sta}) we can see the that a planar orbit is stable, against energy preserving perturbations ($h=0$),  only when the expression between parenthesis is positive  \cite{4,Chandra}.

\bigskip
\noindent{\bf Rotating Lagrangian systems}

Let $q=R(t)Q$, $\dot{R}R^{-1}=\Omega$ then the Lagrangian of a mechanical system 
$L(q,\dq)=\frac{1}{2}\bra\dq,\dq\ket-V(q)$ transforms into

\begin{equation}
L(Q,\dQ) = \frac{1}{2}\bra\dQ,\dQ\ket + \bra\Omega Q,\dQ\ket - V_{ef}(Q)
\end{equation}

\ni in a rotating frame, where $\Omega$ is the rotation matrix and where  the effective potential is 

\begin{equation}
V_{ef}(Q) =V(Q) +\frac{1}{2}\bra \I Q,Q\ket,\quad \I\equiv
\mbox{inertia matrix};
\end{equation}

\ni and $\bra,\ket$ is the usual inner product in $\R^n$.
The prolonged Lagrangian yields the right linearized equations:
\begin{gather}
\ddot{Q}+2\Omega\dQ+ \nabla  V_{ef}(Q)=0\\
\ddot{\epsilon}+2\Omega\dot{\epsilon}+ B\eps=0.
\end{gather}
where $B=HessV_{ef}(Q)$.

\bigskip

\centerline{\textsc{ Acknowledgements.}}
\smallskip

  This work was partially supported by CONACyT and by PAPIIT-UNAM.  ALSB wants to thank C.\ M.\ Arizmendi and G.\ Hentshel for  very interesting conversations in Atlanta. The authors  acknowledge with thanks the useful remarks of G.\ Sardanashvily of Moscow State University and the suggestions of E.\ Pi\~na of UAM-Iztapalapa. Last but not least, HNNY and ALSB dedicate this work to the memory of their beloved friend M.\ Mec.

\end{document}